\documentclass[12pt]{article}
\usepackage{amsfonts}
\usepackage{amssymb}
\usepackage{fancyhdr}
\usepackage{slashed}
\usepackage{graphicx}
\usepackage{subfigure}
\usepackage{color}

\usepackage{hyperref}

\def\hybrid{\topmargin -20pt    \oddsidemargin 0pt
        \headheight 0pt \headsep 0pt
        \textwidth 6.25in       
        \textheight 9.25in       
        \marginparwidth .875in
        \parskip 5pt plus 1pt   \jot = 1.5ex}

\hybrid

\def\baselinestretch{1.2}

\catcode`\@=11

\def\marginnote#1{}
\newcount\hour
\newcount\minute
\newtoks\amorpm
\hour=\time\divide\hour by60
\minute=\time{\multiply\hour by60 \global\advance\minute by-\hour}
\edef\standardtime{{\ifnum\hour<12 \global\amorpm={am}%
        \else\global\amorpm={pm}\advance\hour by-12 \fi
        \ifnum\hour=0 \hour=12 \fi
        \number\hour:\ifnum\minute<10 0\fi\number\minute\the\amorpm}}
\edef\militarytime{\number\hour:\ifnum\minute<10 0\fi\number\minute}

\def\draftlabel#1{{\@bsphack\if@filesw {\let\thepage\relax
   \xdef\@gtempa{\write\@auxout{\string
      \newlabel{#1}{{\@currentlabel}{\thepage}}}}}\@gtempa
   \if@nobreak \ifvmode\nobreak\fi\fi\fi\@esphack}
        \gdef\@eqnlabel{#1}}
\def\@eqnlabel{}
\def\@vacuum{}
\def\draftmarginnote#1{\marginpar{\raggedright\scriptsize\tt#1}}

\def\draft{\oddsidemargin -.5truein
        \def\@oddfoot{\sl preliminary draft \hfil
        \rm\thepage\hfil\sl\today\quad\militarytime}
        \let\@evenfoot\@oddfoot \overfullrule 3pt
        \let\label=\draftlabel
        \let\marginnote=\draftmarginnote
   \def\@eqnnum{(\theequation)\rlap{\kern\marginparsep\tt\@eqnlabel}%
\global\let\@eqnlabel\@vacuum}  }

\def\preprint{\twocolumn\sloppy\flushbottom\parindent 2em
        \leftmargini 2em\leftmarginv .5em\leftmarginvi .5em
        \oddsidemargin -.5in    \evensidemargin -.5in
        \columnsep .4in \footheight 0pt
        \textwidth 10.in        \topmargin  -.4in
        \headheight 12pt \topskip .4in
        \textheight 6.9in \footskip 0pt
        \def\@oddhead{\thepage\hfil\addtocounter{page}{1}\thepage}
        \let\@evenhead\@oddhead \def\@oddfoot{} \def\@evenfoot{} }

\def\numberbysection{\@addtoreset{equation}{section}
        \def\theequation{\thesection.\arabic{equation}}}

\def\underline#1{\relax\ifmmode\@@underline#1\else
        $\@@underline{\hbox{#1}}$\relax\fi}

\def\titlepage{\@restonecolfalse\if@twocolumn\@restonecoltrue\onecolumn
     \else \newpage \fi \thispagestyle{empty}\c@page\z@
        \def\thefootnote{\fnsymbol{footnote}} }

\def\endtitlepage{\if@restonecol\twocolumn \else \newpage \fi
        \def\thefootnote{\arabic{footnote}}
        \setcounter{footnote}{0}}  

\catcode`@=12
\relax

\def\figcap{\section*{Figure Captions\markboth
        {FIGURECAPTIONS}{FIGURECAPTIONS}}\list
        {Figure \arabic{enumi}:\hfill}{\settowidth\labelwidth{Figure
999:}
        \leftmargin\labelwidth
        \advance\leftmargin\labelsep\usecounter{enumi}}}
 \relax
\def\tablecap{\section*{Table Captions\markboth
        {TABLECAPTIONS}{TABLECAPTIONS}}\list
        {Table \arabic{enumi}:\hfill}{\settowidth\labelwidth{Table
999:}
        \leftmargin\labelwidth
        \advance\leftmargin\labelsep\usecounter{enumi}}}
 \relax
\def\reflist{\section*{References\markboth
        {REFLIST}{REFLIST}}\list
        {[\arabic{enumi}]\hfill}{\settowidth\labelwidth{[999]}
        \leftmargin\labelwidth
        \advance\leftmargin\labelsep\usecounter{enumi}}}
 \relax
\makeatletter
\newcounter{pubctr}
\def\publist{\@ifnextchar[{\@publist}{\@@publist}}
\def\@publist[#1]{\list
        {[\arabic{pubctr}]\hfill}{\settowidth\labelwidth{[999]}
        \leftmargin\labelwidth
        \advance\leftmargin\labelsep
        \@nmbrlisttrue\def\@listctr{pubctr}
        \setcounter{pubctr}{#1}\addtocounter{pubctr}{-1}}}
\def\@@publist{\list
        {[\arabic{pubctr}]\hfill}{\settowidth\labelwidth{[999]}
        \leftmargin\labelwidth
        \advance\leftmargin\labelsep
        \@nmbrlisttrue\def\@listctr{pubctr}}}
 \relax
\makeatother
\newskip\humongous \humongous=0pt plus 1000pt minus 1000pt

\newif\ifdtup

\relax

\def\be{\begin{equation}}
\def\ee{\end{equation}}
\def\ba{\begin{eqnarray}}
\def\ea{\end{eqnarray}}

\def\del{\partial}

\def\r{\rho}
\def\a{\alpha}

\def\b{\beta}

\def\g{\gamma}
\def\G{\Gamma}
\def\d{\delta}

\def\th{\theta}

\def\m{\mu}
\def\n{\nu}

\def\l{\lambda}

\def\s{\sigma}

\def\cN{{\cal N}}

 \def\cN{{\cal N}} \def\cO{{\cal O}}

 \def\cZ{{\cal Z}}

\newcommand{\prt}[1]{{\left( {#1} \right)}}

\def\no{\noindent}

\def\qq{\qquad}

\def\IR{\relax{\rm I\kern-.18em R}}

\def \ov {\over}

\def\IR{\relax{\rm I\kern-.18em R}}
\def\IL{\relax{\rm I\kern-.18em L}}

\def\inv{^{\raise.15ex\hbox{${\scriptscriptstyle -}$}\kern-.05em 1}}

\def\bea{\begin{eqnarray}}
\def\eea{\end{eqnarray}}
\newcommand{\eq}[1]{(\ref{#1})}
\def\nn{\nonumber}
\def\del{\partial}
\newcommand{\la}[1]{\label{#1}}

\def\a{\alpha}      
\def\b{\beta}       
\def\g{\gamma}  \def\G{\Gamma}  
\def\d{\delta}

\def\l{\lambda} 
\def\m{\mu} \def\n{\nu}

\def\r{\rho}
\def\s{\sigma}  
\def\t{\tau}
\def\th{\theta}

\def\dxi{\dot \xi}
\def\dt{\dot t}
\def\drho{\dot\rho}
\def\dr{\dot r}

\definecolor{markcolor2}{rgb}{1,0,0}

\definecolor{markcolor3}{rgb}{0,1,0}

\begin{document}

\renewcommand{\theequation}{\thesection.\arabic{equation}}
\csname @addtoreset\endcsname{equation}{section}

\newcommand{\beq}{\begin{equation}}
\newcommand{\eeq}[1]{\label{#1}\end{equation}}
\newcommand{\ber}{\begin{eqnarray}}
\newcommand{\eer}[1]{\label{#1}\end{eqnarray}}
\newcommand{\eqn}[1]{(\ref{#1})}
\begin{titlepage}

\begin{center}


~
\vskip 1 cm

{\large
\bf Non-integrability in non-relativistic theories}

\vskip 0.5in

{\bf Dimitrios Giataganas$^{1,2}$\phantom{x}and\phantom{x}Konstadinos Sfetsos}$^{1,3}$
\vskip 0.1in
{\em
${}^1$ Department of Physics, University of Athens\\
15771 Athens, Greece
\vskip .15in
${}^2$
National Institute for Theoretical Physics\\
School of Physics and Centre for Theoretical Physics\\
University of the Witwatersrand, Wits, 2050, South Africa
\vskip .15in
${}^3$
Department of Mathematics, University of Surrey,\\
Guildford GU2 7XH, UK
\\\vskip .1in
{\tt dgiataganas@phys.uoa.gr, \quad ksfetsos@phys.uoa.gr}\\
}
\vskip .2in
\end{center}

\vskip .4in

\centerline{\bf Abstract}

\no
Generic non-relativistic theories giving rise to non-integrable string
solutions are classified. Our analysis boils down to a simple algebraic
condition for the scaling parameters of the metric. Particular cases are the Lifshitz and the anisotropic Lifshitz spacetimes,
for which we find that for trivial dilaton dependence the only integrable physical
theory is that for $z=1$. For the hyperscaling violation theories we conclude that the vast majority of theories are non-integrable, while only for a small class of physical theories, where the Fermi surfaces belong to, integrability is not excluded.
Schr\"odinger theories are also analyzed and a necessary condition for non-integrability is found. Our analysis is also
applied to cases where the exponential of the dilaton is a monomial of the holographic coordinate.
\end{titlepage}
\vfill
\eject


\noindent


\def\baselinestretch{1.2}
\baselineskip 19 pt
\noindent


\setcounter{equation}{0}

\tableofcontents
\section{Introduction and method}

In recent years there has been intensive research concerning the
integrability of string theoretical models in curved space. The
strings are described by two-dimensional sigma-models and their
classical motion is constrained by the equations of motion which
typically form a non-linear system of differential equations. Among
the plethora of models some are of special interest due to their
integrable structure.

In principle, in order to show that a system is integrable one has to demonstrate that there are as many integrals
of motion as degrees of freedom.
For the cases of sigma-models that we are mostly interested in, the standard way to show integrability is
to find the Lax pair representation,
from which an infinite number of conserved charges follow. In an AdS/CFT context one equivalently derives
a system of integral string Bethe equations which
have been proven equivalent to the thermodynamic limit of the
spin chain Bethe equations. In this way the matching of the multi-spin string energies to the conformal dimension
of the field theory operators is possible \cite{Minahan:2002ve,Beisert:2003yb,Beisert:2010jr}.

The full set of necessary conditions for the existence of a Lax pair
is not known. Considering this, it is natural to think of other ways
to prove or, for that matter, disprove integrability of a system. In
an integrable two-dimensional sigma-model all subsystems that one
can derive by consistent truncations should also be integrable.
Hence, a necessary condition for a two-dimensional sigma-model to be
integrable is that when the system of second order partial
differential equations is consistently reduced to a system of
ordinary differential equations, then this has to constitute an
integrable system. Similarly, one can prove non-integrability by
demonstrating that at least one consistent one-dimensional
truncation of the full system of equations of motion is a
non-integrable system of second-order differential equations. This
can be done either through numerical methods in chaos, or through
semi-analytic algorithmic approaches \cite{Ziglin1,Ziglin2} and
\cite{Fomenko1,Ruiz1,Goriely1}.

These methods of proving non-integrability have been applied so far to cosmological
models as well as in the context of the gauge/gravity correspondence.
Signs of chaotic behavior have been found in Schwarzschild black holes \cite{Frolov:1999pj,Zayas:2010fs}.
In the confining AdS soliton background \cite{Witten:1998zw} the effective Lagrangian of string solutions was found
to be reduced to a set of coupled harmonic and unharmonic oscillators where at high energies
result in a chaotic and non-integrable behavior \cite{Basu:2011dg}.
Moreover, the class of the Sasaki--Einstein manifolds was tested starting with the particular example of the
conifold $AdS_5\times T^{1,1}$ and later for $Y^{p,q}$ manifolds \cite{Basu:2011di,Basu:2011fw}.
The string solutions again show chaotic behavior implying nonintegrability of the corresponding sigma models.
However, in $Y^{p,q}$ manifolds special care should be taken with the string solution to show that indeed they satisfy
the Sasaki-Einstein constraints as in \cite{Giataganas:2009dr,Giataganas:2009an,Giataganas:2010mj}.
More recently it has been shown that integrability is not present in marginal complex
$\beta$-deformations \cite{Giataganas:2013dha}. In addition, working with non-integrable geodesics it has been shown that
integrability is ruled out for $\cN=4$ SYM beyond the planar limit \cite{Chervonyi:2013eja}, although at special
large $N$ limits some integrability does appear \cite{Koch:2011hb,Carlson:2011hy}. Furthermore, the classical
spinning solutions have been found to have chaotic behavior in various confining backgrounds: Klebanov--Strassler,
Maldacena--Nunez, Witten AdS and the AdS soliton \cite{Basu:2012ae}. It is natural to think therefore that confining backgrounds
have a chaotic quantum spectrum in the sense of the Gaussian orthogonal ensemble (GOE) \cite{PandoZayas:2012ig}.
The quantum spectrum has also been studied in the AdS soliton background where a smooth transition from the Wigner GOE
to the integrable regime was noticed as the energy increased, in agreement with the fact that the
system becomes asymptotically $AdS$ at higher energies \cite{Basu:2013uva}.
In \cite{Stepanchuk:2012xi}
the string solutions were studied in curved Dp-brane backgrounds
and non-integrability was found except for special cases. In \cite{Basu:2013vva}
the response of a nonlinearly coupled scalar field in the probe limit of an asymptotically $AdS$ black brane geometry
was studied and it was argued that the response of the dual operator is chaotic. Similarly in \cite{Farahi:2014lta}, the response of the scalar field to the gravitational collapse in an asymptotically Anti de Sitter
space-time  was found to be chaotic.

In this paper we extend these studies to the non-relativistic
gauge/gravity dualities. There are many interesting systems which have general dynamical scalings instead of scale invariance,
as for example the Lifshitz, the spatial anisotropic Lifshitz-like and the Schr\"odinger spaces,
as well as backgrounds with hyperscaling violations of the above.
General Lifshitz geometries may be constructed, in a phenomenological way,  by switching on fluxes with non-trivial topological
couplings \cite{Kachru:2008yh}. Explicit Lifshitz solutions in string theory are more difficult to construct.
Progress in that direction has been made in \cite{Hartnoll:2009ns,Balasubramanian:2010uk} and in \cite{Donos:2010tu} where it has been proved that
Lifshitz solutions with dynamical
exponent $z=2$ can be embedded in supergravity. Later in \cite{Gregory:2010gx} explicit type-II supergravity
solutions of the form $Lif_p \times \Omega_{d-p}$ with generic dynamical exponents, where $\Omega_{d-p}$
is a constant curvature spherical flat or hyperbolic space, have been found. In the meantime anisotropic
Lifshitz-like solutions where found in \cite{Azeyanagi:2009pr} and in addition they have been extended to finite
temperature cases \cite{Mateos:2011tv} and also used to study anisotropic plasmas,
eg. \cite{Giataganas:2012zy, Giataganas:2013lga}.
The non-relativistic conformal field theories can also be constructed as cosets of the Schr\"odinger group
and its variants as in \cite{SchaferNameki:2009xr}. Moreover, it has been found that the IR geometry of a dilaton theory with an asymptotic electric flux is Lifshitz \cite{Goldstein:2009cv}. Other non-relativistic constructions have been found in \cite{Narayan:2012hk}.

Motivated by all these developments we consider the integrability of string solutions in non-relativistic theories.
 We find generic conditions in terms of the scaling parameters of the metric elements of the backgrounds,
 which when not satisfied imply non-integrability.
For the particular example of the Lifshitz backgrounds with trivial dilaton dependence we find that the only integrable solution is the
case with $z=1$, corresponding to the $AdS$ space. When the hyperscaling violation scaling is considered, only for a subset of all possible backgrounds integrability is not excluded. This subset includes the Fermi surfaces. We derive explicitly the bounds of the non-integrable physical theories.  We additionally analyze spaces with non-diagonal metric terms, where the Schrodinger metrics
 are particular examples and we find cases that integrability is excluded.

It is interesting that for the cases we find non-integrability,
there is no sign of chaos inside the accepted range of the
parameters and coordinates and when satisfying the Virasoro
constraints. We checked this for solutions obtained from a wide
choice of the initial values in the system of differential
equations. Notice that it does not necessarily mean that for other
energies, not fixed by the Virasoro constrain, or other string
configurations the system is not chaotic. The integrability of dynamical
systems is related on how the system reacts to small variations
around its phase space curves. According to the
Kolmogorov--Arnold--Moser (KAM) theorem, when the system is
non-integrable it is resonant in the corresponding phase space
describing the motion in the angle variables. Practically, when weak
non-linear perturbations are applied to an integrable Hamiltonian
system whose motion is confined to invariant tori, then some
deformed invariant tori may still be present and some others may be
destroyed. One would like to know, whether or not the perturbed
solutions are stable and if the perturbed orbits will remain close
to unperturbed ones for long period of time. The conditions for the
regular and normal behavior of the perturbed solution are provided
by the KAM theorem. Chaos can occur only when the KAM theorem does
not hold. Chaotic motion arises due to some appropriate noise
applied to usually well behaved solutions, resulting to a motion
that is highly sensitive to initial conditions. Moreover, the
appearance of chaos can also be viewed as a breakdown of
integrability, but non-integrability does not always imply chaotic
behavior.

The chaotic motion can be realized easier from the Poincare maps. By fixing the energy of the solutions of the Hamiltonian system
and choosing one invariant plane, one can obtain curves on a two dimensional plane generated by fixing the initial conditions.
When the curves are in a sense continuous, the solutions are regular.  These
regular trajectories are obtained by integrating the equations of system for the particular conditions, and
 the solutions are unique and reproducible. When individual points appear, the chaotic regions become
apparent. These points are located randomly
and integrability breaks down over the phase space. There is a possibility to have
regions of phase space which show integrability and regions
of chaos for different initial conditions. For our solutions we have not found regions of chaotic motion for
the particular fixed energy of the system
imposed by the zero Hamiltonian. This does not mean that the system is not chaotic, since in other energies chaos may occur.

Alternatively, to the aforementioned techniques, by using
differential Galois theory, it can be examined if the identity
component of the differential Galois group of variational equations
normal to an integrable plane of solutions is Abelian. However, in
practice, in order to determine the Galois group is a difficult
task. Therefore, one takes advantage of the fact that the Abelian
identity element in the Galois group is equivalent to finding
Liouvillian solutions, i.e. combinations of algebraic functions,
exponentiation of quadratures and quadratures, to the normal
variational equation. This is in principle an easier task, and for
the Hamiltonian systems we are interested in, we can end up with
second order homogeneous differential equations with rational
coefficients, where the existence of the Liouvillian solutions, can
be checked with the running of the Kovacic algorithm
\cite{Kovacic1}, which only fails when the system has no Liouvillian
solutions.

The plan of the paper is as follows: In the section 2 we introduce the theories of special interest
we plan to examine as particular cases of the generic analysis. In section 3,  we start by working with a generic sigma model.
Then we study the integrability of string configurations in general non-relativistic spaces and apply
our results to backgrounds of special interest. In section 4, we derive the conditions of non-integrability
for metrics with non-diagonal terms and apply them to known backgrounds.
In section 5, we conclude and discuss our results and further directions.

\setcounter{equation}{0}

\section{Classes of theories we focus on}

In this section we give a brief overview of the properties of a class of backgrounds of special interest which are particular examples of the generic study in the subsequent sections. 

Let us a consider $(d+2)$-dimensional Lifshitz metrics of the form
\be
\la{m2}
ds^2=r^{- 2\theta/d}\left(-r^{2z}dt^2+r^2 \sum_{i=1}^d dx_i^2 + \frac{dr^2}{r^2}\right)\ ,
\ee
where the constants $z$ and $\th$ are the dynamical critical and hyperscaling violation exponents, respectively.
This metric transforms covariantly under the scaling transformation
\be
\quad t\rightarrow \lambda^z t\ ,\quad x_i\rightarrow \lambda x_i\ ,\quad r\rightarrow r/\lambda\ ,\quad  ds\rightarrow \lambda^{\th/d}ds\ .
\ee
In the $SO\prt{d}$-invariant plane spanned by the coordinates $x_i$, $i=1,2,\dots ,d$, we single out the
 $1\!-\! 2$ plane in which we use polar coordinates
($\r,\phi)$.

We plan to study the non-integrability of such holographic theories, although our results may apply widely in
theories of general gravitational interest.
In addition, we will consider the restrictions imposed in the present context by the null energy condition
\be
T_{\m\n}N^\m N^\n\geqslant 0\ ,
\label{tmn}
\ee
in order for the dual field theory to be physically sensible \cite{Dong:2012se}.
This implies that
\be
\la{cond1}
\prt{d-\th}\prt{dz-\th-d}\geqslant 0\ ,\qquad \prt{z-1}\prt{d+z-\th}\geqslant 0 \ .
\ee
For $\th=0$ we recover the Lifshitz space $Lif_z$ physical constraint with the condition $z\geqslant 1$.
The saturation of the bound happens for $z=1$ where the spacetime becomes $AdS_{d+2}$.
The exponent $\th$ is holographically related to the thermal entropy density in
the dual theory which scales as $S\simeq T^{\prt{d-\th}/z}$.

An additional physical requirement for any local quantum field theory is that the entanglement entropy of a given region scales with the area of its boundary up to logarithmic corrections.
This gives the condition \cite{Huijse:2011ef}
\be
\la{cond2}
\th\leqslant d-1\ ,
\ee
which also excludes non-stable theories. It has been shown that when this bound is saturated for $\th=d-1$, the dual theory exhibits a Fermi surface \cite{Ogawa:2011bz,Huijse:2011ef}.
Then the condition \eq{cond1} simplifies to $z\geqslant 2-1/d$. When this
is also saturated, the corresponding values of the scaling parameters are of particular interest,
since they can be related to the non-Fermi liquid states, for example the $d=2$, giving $\th=1$ and $z=3/2$ \cite{Ogawa:2011bz}.

Our discussion can be generalized to anisotropic Lifshitz-like theories. For vanishing hyperscaling parameter
$\th$ the corresponding metric takes the form
\be
\la{llikem}
ds^2=r^{2z}\left(-dt^2 + \sum_{i=1}^k dy_i^2\right) +r^2\sum_{i=k+1}^d dx_i^2+\frac{dr^2}{r^2}\ ,
\ee
for which the $SO(d)$ symmetry of the spatial part is broken down to $SO(k)\times SO(d-k)$.
In this case the parameter $z$ measures not only the degree of Lorentz symmetry violation, but also the anisotropy of the full space
spanned by all the $y_i$'s and $x_i$'s. This metric is invariant under the scaling
\be
t\to \l^z t\ ,\quad y_i\to \l^z y_i\ ,\quad x_i\to \l x_i \ ,\quad r\to r/\l\ .
\ee
The non-relativistic metrics we are interested in may also have a non-trivial $r$-dependent dilaton field $\Phi$
when embedded in the string theory,
i.e. in practice in type-II supergravity.
In those cases the factor $\displaystyle e^{\Phi}$ may have a monomial form, as it happens in many of the known supergravity solutions.

Another class of non-relativistic theories we will study correspond to the Schr\"odinger spacetime $Sch_z$,
for which the metric can be written as
\be\la{schrm}
ds^2=-r^{2z}dt^2 +  r^2 \prt{2 dt d\xi+d\rho^2+\rho^2 d\phi^2+\sum_{i=3}^d dx_i^2}+\frac{dr^2}{r^2}\ .
\ee
The isometry algebra of the $Sch_z$ spaces contains the Galilean boosts as well as null translations of $\xi$.
The latter play the r\^ole of the central extension or mass operator of the Galilean algebra.
The presence of the off-diagonal term of the $Sch_z$ space modifies the details of the approach for
 showing non-integrability compared to that for the $Lif_z$ case.
Hence we will analyze first the non-integrability of $Lif_z$ spacetime and then that of the $Sch_z$ one.

\section{Non-Integrability in generic non-relativistic spaces}
\label{section:gen}

In this section we first develop a general formalism and then we focus on a more applied analysis to the cases we have mentioned.

\subsection{Consistency of sigma-model solutions}

We first discuss general properties of the string solutions we construct after we transform to the string frame metric. These are localized in all directions except for a single cyclic coordinate which we allow to wind.
We will show that our ansatz is consistent and that the Virasoro constraints do not contribute additional
differential conditions to the system. They just constrain the allowed initial conditions.

Consider a general two-dimensional sigma-models of the form
\be
S(X) = - {\sqrt{\l}\ov \pi} \int Q_{\m\n} \del_+ X^\m \del_- X^\n\ ,\qq Q_{\m\n}= G_{\m\n}+ B_{\m\n}\ .
\label{eq1}
\ee
We use the coordinate conventions $ x^{\pm } = \tau \pm \s$.
The equation of motion for $X^\m$  is
\be
\ddot X^\m - X^{''\m} + \G^\m{}_{\n\r}(\dot X^\n \dot X^\r - X^{'\n}X^{' \r}) + H^\m{}_{\n\r} \dot X^\n X^{'\r} = 0 \,,
\ee
where $H_{\r\m\n}=\del_\r B_{\m\n}+\del_\n B_{\r\m}+\del_\m B_{\n\r}$. The Virasoro constraints are given by
\ba
&& G_{\m\n}(\dot X^\m\dot X^\n + X^{'\m} X^{' \n}) = 0 \ ,
\nonumber\\
&& G_{\m\n}\dot X^\m  X^{\prime \n}=0 \ .
\ea
We will assume that there is a Killing isometry corresponding to constant shifts of the coordinate $X^9$.
We will let accordingly, the rest of the coordinates
be denoted by $X^i$, so that the index $\m=(i,9)$.
Consider next solutions of the form
\be
\la{ann1}
 X^i = X^i(\tau)\ , \qq X^9 = m \s\ , \quad m \in \mathbb{Z}\ .
\ee
Then the equations of motion become
\ba
&&\m= i: \qquad  \ddot X^i + \G^i{}_{jk}\dot X^j \dot X^k - m^2 \G^i{}_{99} + m H^i{}_{j9}\dot X^j = 0 \ ,
\nonumber
\\
&&\m= 9: \qquad \G^9{}_{ij} \dot X^i \dot X^j -m^2 \G^9{}_{99} + m H^9_{i9}\dot X^i = 0 \ .
\label{dkjg3}
\ea
Similarly, the Virasoro constraints become
\ba
&& G_{ij}\dot X^i \dot X^j  +m^2 G_{99} = 0\ ,
\nonumber\\
&& m G_{9i} \dot X^i = 0 \ .
\label{fhk2}
\ea
By taking the $\tau $ derivative of the first Virasoro constrain and using the equations of motion we show that the constrain is preserved and therefore one just needs to satisfy them by imposing appropriate initial conditions.
In addition, we assume that
\be
\la{condd1}
G_{9i}=0\ .
\ee
Then, the second equation of \eq{dkjg3} is satisfied identically.

\no
Therefore, for a theory satisfying \eq{condd1} and for the particular string configuration \eq{ann1}, we are left
only with the first equation of motion \eq{dkjg3}.
An effective one-dimensional action from which this non-trivial equation can be derived from, is given by
\be
S_{\rm 1\!-\!dim} = - \frac{\sqrt{\l}}{2} \int d\tau (G_{ij}\dot X^i \dot X^j - m^2 G_{99} + 2 m B_{9i} \dot X^i)\ .
\label{1dims}
\ee
The Hamiltonian and conjugate momentum are given by
\be
H_{\rm 1\!-\! dim} = - {\sqrt{\l}\ov 2} (G_{ij} \dot X^i \dot X^j + m^2 G_{99})=0\ ,\qq P_i = - G_{ij} \dot X^j  - m B_{9i}\ ,
\label{1dimh}
\ee
where setting $H_{\rm 1-dim}=0$ implements the Virasoro constraint as in \eqn{fhk2}. In the next sections we consider supergravity solutions with
zero B-fields. In several cases, our analysis can be applied even when B-fields are present. This is when the string configuration can be localized consistently such that no new contributions appear from the extra field. We elaborate further on that in the case of Schr\"{o}dinger spaces.

\subsection{Non-integrability of the string solutions}\label{section:sgen}

The set of backgrounds we will consider in this section include as special cases, the hyperscaling violation metrics,
as well as Lifshitz and Lifshitz-like backgrounds with  possible non-trivial dilaton dependence.
The metric
\be
\la{gen1}
ds^2=-r^\a dt^2+r^\b\prt{d\rho^2+\rho^2 d\phi^2+\sum_{i=3}^d dx_i^2}+r^\g dr^2\ ,
\ee
captures all of these cases by appropriately choosing the parameters $\a,~\b$ and $\g$ .

We consider an extended string along a $U(1)$ isometry corresponding to $\phi$, i.e. $\phi=m\s$ and also take the $x_i$'s in the space \eq{gen1} to be constant which obviously in consistent
with the equations of motion. For the variables $t$, $r$ and $\r$ we will allow only $\tau$ dependence.
Using the results of the previous section, where the metric element $G_{99}:=G_{\phi \phi}$, we find that the effective one-dimensional action
\eqn{1dims} of the system is given by
\be
S_{\rm 1\!-\! dim}= - {\sqrt{\l}\ov 2} \int d\t \left(-r^{\a} \dot t^2 + r^{\g}\dr^2 + r^{\b}(\drho^2 - m^2 \rho^2)\right)\ .
\label{dskhk3}
\ee
The corresponding equations of motions are
\bea
&&t:\quad \del_\tau\prt{\dt r^{\a}}=0~\ ,
\nn\\
&&\rho:\quad \del_\tau \prt{\drho r^{\b}} + m^2 r^{\b} \r =0~\ ,
\la{eomr12}\\
&&r: \quad  \del_\tau\prt{2 r^{\g}\dr } + \a r^{\a-1} \dt^2-\g r^{\g-1}\dr^2 - \b r^{\b-1}(\drho^2 - m^2 \r^2)=0
\nn
\eea
and the vanishing Hamiltonian implementing the Virasoro constraints is
\be
H_{\rm 1\!-\! dim} = - {\sqrt{\l}\ov 2} \left[-r^{\a} \dt^2+r^{\g}\dr^2+r^{\b}\prt{\drho^2+ m^2\rho^2}\right]=0\ .
\label{hpart2}
\ee
The equation of motion for $t$ in \eq{eomr12} is solved for
\be
\la{tsol}
\dt=\frac{c}{r^{\a}} \ ,
\label{tsol2}
\ee
where $c$ is the integration constant which can be taken non-negative with no loss of generality.
Then, the remaining equations in \eq{eomr12} become
\bea
&&\ddot{\r}+\b \frac{ \dot{r} }{r} \dot{\rho}+ m^2 \rho =0\  ,
\nn\\
&& c^2 \a r^{-\a-1}+\frac{1}{\dr}\partial_\tau\prt{r^\g \dr^2}+\b r^{\b-1}\prt{-\drho^2+\r^2 m^2}=0 \ .
\label{sy1}
\eea
These equations can be derived from the effective one-dimensional action for the fields $\r$ and $r$, given by
\be
S_{\rm eff}= - {\sqrt{\l}\ov 2} \int d\tau \left( r^\g \dr^2+r^\b\prt{\drho^2-m^2\r^2 }+ c^2 r^{-\a}\right)\ .
\label{js33}
\ee
The corresponding Hamiltonian is\footnote{Note that $H_{\rm eff}$ below coincides with $H_{\rm 1\!-\!dim}$ in \eqn{hpart2}
after we substitute the solution \eqn{tsol2}. This is not the case at the level of the corresponding Lagrangians in \eqn{dskhk3}
and in \eqn{js33}.}
\be
H_{\rm eff}=- {\sqrt{\l}\ov 2} \left(r^\g \dr^2 + r^\b\prt{\drho^2+\r^2 m^2} - c^2 r^{-\a} \right)\ .
\ee

Hence, we have consistently reduced the original string system to a classical one-dimensional Hamiltonian system for two fields
$\r$ and $r$.
A solution to the first equation in \eqn{sy1} is obtained for
\be
\rho=0\ .
\label{sgjsol}
\ee
Then, the equation for $r$ and its solution are given by
\be
\la{req1}
\dr^2=c^2 r^{-\a-\g}\quad \Longrightarrow\quad r(\t) = \left({c\over 2} (\a+\g +2) \tau\right)^{2\over \a+\g +2}\ ,
\label{classr}
\ee
where, with no loss of generality, we have set a constant of integration to zero. The solution is valid if $\a+\g+2\neq 0$.
In the opposite case we comment appropriately below.

To obtain the NVE we perturb in the transverse plane so that $\rho=0+\eta\prt{t}$, obtaining easily from the
first of \eqn{sy1} that\footnote{One could also perturb the classical solution for $r(\tau)$ in \eqn{classr}.
It should be obvious from the form of the system in \eqn{sy1} that the two fluctuations do not couple to linear order and therefore
we may ignore the perturbation of $r$.}
\be
\la{nvegen}
\ddot{\eta}+ \b \frac{\dr}{r}\dot{\eta}+ m^2 \eta =0\ .
\ee
Using \eqn{req1} to substitute for $\dot r/r$, we obtain that
\be
\ddot{\eta}+ \frac{1-2\n }{\t} \dot{\eta} + m^2 \eta=0  \ ,\qq \n = {\frac{1}{2}-\frac{\b}{\a+\g+2}}\ ,
\label{nve00}
\ee
which has rational coefficients. Its general solution is given by
\be
\la{nvea1}
\eta\prt{\t}=\t^\n \left(c_1 J_\n(m \t) + c_2 Y_\n(m \t)\right) \ ,
\ee
where $J_\n$ and $Y_\n$ are the Bessel functions of the first and of the second kind, respectively.

It is known that only when the order $\n$ of the Bessel functions $J_\n(x)$ and $Y_\n(x)$ is
half integer, then they are expressible in terms of elementary functions $\sin x$ and $\cos x$ divided by positive integer powers of $\sqrt{x}$.
In that case, the NVE \eq{nvea1} admits Liouvillian solutions, the string solutions are integrable and the corresponding theory could be
integrable. In all the other
cases the Hamiltonian system, and the corresponding background is non-integrable.
More explicitly the Liouvillian solutions appear for
\be\la{z0}
\n=\frac{2j+1}{2} \qquad \Longrightarrow \qquad \frac{2+\a+\g}{\b} =-\frac{1}{j}
\ , \quad j\in \cZ\ .
\ee
Therefore, when \eq{z0} is not satisfied for the metric form of \eq{gen1} in string frame then the theory is not integrable.
Notice that for $j=0$, implying $\b=0$, \eqn{nve00} becomes the harmonic oscillator equation
and the string configuration is integrable, independently of the value of $\a$ and $\g$.

\no
Turning to the case with $\a+\g+2=0$, the solution of \eqn{req1} is simply $r = e^{c\tau}$.
Then \eqn{nve00} becomes the harmonic oscillator with friction equation with solutions given by elementary functions.
Hence the corresponding string is
also integrable.

\no
As a side remark notice that for the case of no string winding, i.e. $m=0$, the resulting geodesics have always
integrable structure.\footnote{When $m=0$, the solution of \eqn{nve00} is a linear combination of a $1$ and $\tau^{2\n}$.}
This is in agreement with the vast majority of the works in the literature, since
non-integrability is more likely to be probed by extended strings.


\subsection{Application to theories of special interest}
\la{sec:special}

\subsubsection{Lifshitz non-integrability}

Consider the Lifshitz metric
\be
\la{lm1}
ds^2=-r^{2z}dt^2+r^2\prt{d\rho^2+\rho^2
d\phi^2+\sum_{i=3}^d dx_i^2}+\frac{dr^2}{r^2} \ ,
\ee
with a trivial dilaton. This is a particular case of the metric \eq{gen1} with
\be
\a=2 z\ , \quad \b =2 \ ,\quad \g=-2\ .
\ee
Then the condition \eqn{z0} gives
\be
\la{z2}
 z =-\frac{1}{j}  \ , \quad  j\in \cZ\ .
\ee
These are the only values for which we may have integrability.
Specifically, the exact values, of $z$ for which the Hamiltonian system is integrable are when
\be
\label{setz}
z\in \left\{-1,-\frac{1}{2},-\frac{1}{3},-\frac{1}{4},\cdots , \frac{1}{4},\frac{1}{3},\frac{1}{2},1\right\} \ .
\ee
The null energy condition is satisfied for $z\geqslant 1$. Therefore the only physical Lifshitz theory that is
integrable is for $z=1$, i.e. the $AdS$ background.
Backgrounds with $z<1$, where the dual field theories are not well defined, might be integrable for the values of $z$ defined in \eq{z2}.

Moreover, there are several interesting Lifshitz backgrounds in the literature with dilatons depending
on the holographic coordinate $r$. For example, it has been found that the $AdS_5\times S^5$
geometries backreacted by an $\cO\prt{N^2}$ density of heavy quarks, exhibit a Lifshitz geometry
in the infrared with a dynamical critical exponent $z=7$ and a dilaton having a $\ln r^6$
dependence on the holographic coordinate \cite{Kumar:2012ui}. Applying the equation \eq{z0} we obtain $j=-1/4$
and therefore the theory is not integrable at least in the IR. The non-integrability of the theory may have even
further implications to the observables of the theory, for example chaos may appear in some of them, and this worth a further investigation.

\no

\subsubsection{Anisotropic Lifshitz-Like non-integrability}

It is not difficult to see that it is consistent to localize the string along the $x_i$ directions of the anisotropic Lifshitz-like metric \eq{llikem} by setting all the $x_i$'s to constant values.
Then we may follow an identical analysis to the one corresponding to the metric \eqn{lm1} leading eventually to the
conclusion that if \eqn{setz} is satisfied then the corresponding string model may be integrable. For all other values integrability is
excluded. Moreover, from the null energy condition the
only physically allowed value allowing integrability is $z=1$, corresponding to the $AdS$ background.

A particular such anisotropic background is the holographic dual of
Lifshitz-like fixed point found in type-IIB
 supergravity in \cite{Azeyanagi:2009pr}, where a non-trivial dilaton exist. In the Einstein frame the metric
is equivalent to \eq{llikem} with $z=3/2$, $k=2$ and $d=3$, while it
has a non-trivial dilaton with $\ln r^{2/3}$ dependence. In the string frame it becomes
equivalent to the metric \eq{gen1} with $\a=\b=7/3$ and $\g=5/3$ and the
remaining anisotropic directions scale with a factor $r^\g$. However,
they play no r\^ole in the analysis since the string is
localized in there. Using the condition \eq{z2} we get the non-integer value $\n=8/7 \notin \cZ$ and therefore the theory is not integrable.

\subsubsection{Hyperscaling violation metrics non-integrability}

Comparing the metrics \eq{m2} and \eqn{gen1} we identify the parameters
\be
\la{hyp1}
a=2z -2\frac{\theta}{d}\ ,\qquad \b=2-2\frac{\theta}{d}\ ,\qquad\g=-2-2\frac{\theta}{d}\ .
\ee
Using \eq{z0} we find that the corresponding NVE has Liouvillian solutions for the values
\be
\la{z1}
\frac{\th}{d}=\frac{1+j z}{1+2 j} \ , \quad j\in \cZ\ .
\ee
For any other values of $\th$ and $z$, not satisfying the above conditions the corresponding theory is not integrable. 
By applying the null energy condition \eq{cond1} and the one obtained from entanglement entropy scaling \eq{cond2}
to ensure for physical dual gauge theories, we find that the necessary conditions for integrable physical theories to exist are
\be
\la{conhyp}
j=-1\ ,\qq \th=d\prt{z-1}\ ,\qquad 1\leqslant z\leqslant 2-\frac{1}{d}\ .
\ee
For all other values of $d,~z$ and $\th$ the theory is not integrable.
The lower value $z=1$ corresponds to the $AdS_{d+2}$ case and the upper one $z=2-1/d$ to
gravity dual theories for  Fermi surfaces. The saturation
of these conditions happens among other values for $d=2,~\th=1,~z=3/2$ which occur for non-Fermi liquid
states \cite{Ogawa:2011bz} and correspond to regular IR limit. We conclude that the majority of physical and non-physical theories is non-integrable\footnote{We do not consider negative values of the parameter $\th$ or values for $d-1<\th<d$. In case that the constraints of the physical theories will be relaxed to include some of these values and we end up with a slightly modified condition \eq{conhyp}.}.

\section{Geometries with non-diagonal terms}

In this section we consider metrics of the form
\be\la{nond}
ds^2=-r^\a dt^2 + 2r^\d dt d\xi+r^\b\prt{d\r^2+\r^2d\phi^2 + \sum_{i=3}^d dx_i^2}+r^\g dr^2\ ,
\ee
which is similar to \eqn{gen1}, but with the extra off-diagonal term.
In this class of backgrounds the Schr\"odinger geometries with trivial or non-trivial polynomial $r$-dependent
 dilaton are included as particular cases.

\no
Proceeding similarly to section 3 we  set the $x_i$'s to constants, $\phi=m\s$ and then we obtain  the
effective one-dimensional action
\be
S_{\rm 1\!-\!dim}=- {\sqrt{\l}\ov 2} \int d\t \left(-r^{\a} \dot t^2 + r^{\g}\dr^2 + r^{\b}(\drho^2 - m^2 \rho^2) + 2 r^\d \dot t \dot \xi \right)\ ,
\label{dskhk3xi}
\ee
for the remaining variables.
The corresponding equations of motions are
\bea
&&t:\quad \del_\tau\prt{\dt r^{\a} - r^\d\dot \xi}=0\ ,
\nn\\
&&\xi:\quad\del_\tau\prt{r^\d\dt}=0\ ,
\nonumber\\
&&\rho:\quad \del_\tau \prt{\drho r^{\b}} + m^2 r^{\b} \r =0~\ ,
\la{eomr12xi}\\
&&r: \quad  \del_\tau\prt{2 r^{\g} \dr } + \a r^{\a-1} \dt^2-\g r^{\g-1}\dr^2 - \b r^{\b-1}(\drho^2 - m^2 \r^2) - 2 \d r^{\d-1} \dt\dxi =0
\nn
\eea
and the vanishing Hamiltonian implementing the Virasoro constraint is
\be
H_{\rm 1\!-\!dim} =- {\sqrt{\l}\ov 2} \left[-r^{\a} \dt^2+r^{\g}\dr^2+r^{\b}\prt{\drho^2+ m^2\rho^2} + 2  r^{\d} \dt\dxi \right]=0\ .
\label{hpart2xi}
\ee
The first of the equations in \eqn{eomr12xi} are easily integrated once to give
\be
\dt=\frac{c}{r^\d}\ ,\qquad \dxi={c\ov r^{2\d-\a}} - \frac{c_1}{r^\d}\ ,
\ee
where $c$ and $c_1$ are the integration constants.
Then the remaining equations become
\bea
&&\ddot{\r}+\b \frac{ \dot{r} }{r} \dot{\rho} + m^2 \rho =0\  ,
\nn\\
&& \frac{1}{\dr}\partial_\tau\prt{r^\g \dr^2}+\b r^{\b-1}\prt{-\drho^2+m^2 \r^2} + c^2 (\a-2\d) r^{\a -2\d-1}   +2 \d c c_1 r^{-\d-1} =0 \ .
\label{sy1xi}
\eea
These equations can be derived from the effective one-dimensional action given by
\be
S_{\rm eff}= -{\sqrt{\l}\ov 2} \int d\tau \left( r^\g \dr^2+r^\b\prt{\drho^2-m^2\r^2 } -  c^2 r^{\a-2\d} +2 c c_1 r^{-\d}\right)\ .
\label{js33xi}
\ee
The corresponding Hamiltonian is
\be
H_{\rm eff}=- {\sqrt{\l}\ov 2}\left(r^\g \dr^2 + r^\b\prt{\drho^2+\r^2 m^2} + c^2 r^{\a-2\d} -2 c c_1 r^{-\d} \right)\ .
\ee
Hence, we have consistently reduced the original string system to a classical one-dimensional Hamiltonian system for two fields
$\r$ and $r$.

Choosing the same solution as in \eqn{sgjsol} we find that
\be
\dr^2=c r^{-\d-\g}\prt{2 c_1-c r^{\a-\d}}\ ,
\label{dsagxi}
\ee
which for generic values of $\a$, $\b$ and $\d$ has solution in terms of inverse hypergeometric functions.
By varying the equation of motion for $r$ with $\rho=0+\eta\prt{t}$ we obtain  the same form of the equation as in \eqn{nvegen}.
Then using \eqn{dsagxi} we obtain
\be
\la{snve}
\eta'' + \prt{\frac{2 \b-\d-\g}{2 r}+\frac{c\prt{\d-\a}}{2\prt{2 c_1 r^{\d-\a+1}-c r} }}\eta'+ m^2 \eta=0\ ,
\ee
where the derivative now is with respect to $r$.
The NVE has rational coefficients when $\d-\a+1 \in \cZ$, then the Kovacic algorithm can be applied.
In contrary to the diagonal metrics, we do not find for generic values of $\a,\b,\g$ and $\d$ analytic solutions,
so that we cannot tell when they become Liouvillian. Therefore to show non-integrability, we need to apply particular values to the
parameters of our system, directly to the above NVE, as we do in the case of the Schr\"{o}dinger geometries below.
Notice that by localizing the string, i.e. choosing $m=0$, we see that any geodesic of the type we have chosen is integrable,
since the solutions to the above NVE are integrals of polynomials and square roots.

\subsection{Application to Schr\"{o}dinger geometries}
\la{sec:special2}

The $Sch_z$ geometries are obtained from the metric \eq{nond} by setting
\be
\a=2z~,\quad\d=2~,\quad \b=2~,\quad\g=-2~.
\ee
We find that for the values $z= 1, 2, 3$, 
then \eq{snve} has Liouvillian solutions, the corresponding strings are integrable and we can not exclude integrability of the theory.
For $z=4,5,6$ we have non-integrable systems. There are also higher values of $z$ where integrability can
not be excluded. For any value of $z$ the test of integrability can be done directly by checking the type of
solutions of \eq{snve}. We find that for the vast majority of the $z$ values our strings are integrable.

Notice that the $z=2$ Schr\"{o}dinger spaces can be constructed with
an appropriate TsT transformation \cite{Herzog:2008wg}  and has been
shown to be integrable \cite{Orlando:2010yh,Kawaguchi:2011wt} which is in agreement
with our results. Indeed, our analysis can be applied directly to
these backgrounds by localizing consistently the string
configuration such that there is no contributions to the effective
action by the antisymmetric tensor field which is present in these
backgrounds.

\section{Conclusions}

In this paper we have used methods of classical integrability to study non-relativistic theories.
Working in a general framework, our analysis boils down to a simple formula of the scalings of the metric
 elements which when is not satisfied integrability is excluded. This is achieved by analyzing some unique
properties of the solutions of the differential NVE  equations which are obtained from string configurations
 in spaces under study. As a next step we have applied our results to non-relativistic theories of special interest.
In particular the physical Lifshitz theories with trivial dilatons are integrable only for the value $z=1$ corresponding to the $AdS$ geometry.
Thus any other candidate theory of different scaling is excluded in terms of integrability.  The same applies to
 the Lifshitz-like theories since the non-integrable string solutions we have found can be localized consistently
in the additional anisotropic directions. In theories with
hyperscaling violation scaling we have found that in the small range
of physical theories for which integrability is not excluded, the
Fermi-surface theories saturating the null energy condition, are
included. We also performed the analysis for non-relativistic
theories with non-diagonal metric elements. For these the study of
integrability needed to be done case by case due to the absence of
similar unique properties of the solutions of the corresponding
NVEs.

For theories containing non-trivial dilaton, we brought their
metrics to the string frame and applied our generic formulas. As an
example, we explicitly checked the type-IIB anisotropic
Lifshitz-like supergravity solutions of \cite{Azeyanagi:2009pr} and
we found that it is not integrable.

For vanishing winding number, all of our geodesics are integrable.
This means that in order to see the non-integrable structure of the
model we need to probe it with extended strings. This is a common
finding when using methods of classical integrability in the string
theory models and is in some sense expected, since the richer
structure of the string reveals the non-integrable structure of the
theory. Note also that, in general, a chaotic motion implies
non-integrable solutions but the opposite is not always true. In our
solutions with the constraints set by the zero Hamiltonian,  we
detect no chaotic behavior, in the physical range of the parameters
and coordinates. This finding does not necessarily exclude chaos in other range
of parameters and energies or for other string solutions.

We also note that by excluding integrability in the non-relativistic spaces we have worked, we expect
non-integrability to follow for a wide range of spaces that have as a particular limit these non-relativistic spaces. For example,
for the geometries backreacted by a heavy quark density which in the IR limit flow to a Lifshitz geometry, we expect non-integrability.

It would be very interesting to investigate the implications of our findings to the observables of the dual NRCFT.
 We have found non-integrability in backgrounds which are used to describe dual anisotropic plasmas, where a wide range of
observables has been already  investigated \cite{Giataganas:2013lga}, backgrounds with heavy quark density and backgrounds
used to describe condensed matter systems. Certain predictions of these non-integrable theories may be chaotic,
like the hadronic spectrum in confining theories \cite{PandoZayas:2012ig}.
In our case would be interesting to investigate this possibility and the physical implications of the chaos to the observables.

\newpage
\subsection*{Acknowledgments}

We are thankful to P. Kumar, K. Siampos, A. Tseytlin and K. Zoubos  for useful conversations, comments and correspondence.
The research of the authors is implemented under the ``ARISTEIA'' action of the ``operational programme education and lifelong learning''
and is co-funded by the European Social Fund (ESF) and National Resources.

\bibliographystyle{JHEP}

\end{document}